\documentclass[preprint,showpacs,preprintnumbers,amsmath,amssymb]{revtex4}

\usepackage{graphicx}
\input{epsf}

\begin{document}

\title{Conductivity of 2D many-component electron gas,  partially-quantized by
magnetic field}

\author{M.V.Entin}
\affiliation{Institute of Semiconductor Physics, Siberian Branch
of Russian Academy of Sciences, Novosibirsk, 630090, Russia}
\author{L.I.Magarill}
\affiliation{Institute of Semiconductor Physics, Siberian Branch
of Russian Academy of Sciences, Novosibirsk, 630090, Russia}
\begin{abstract}
The 2D semimetal  consisting of heavy holes and light electrons is
studied. The consideration is based on assumption that electrons
are quantized by magnetic field while holes remain classical. We
assume also that the interaction between components is  weak and
the conversion between components is absent. The kinetic equation
for holes colliding with quantized electrons is utilized. It has
been stated that the inter-component friction and corresponding
correction to the dissipative conductivity $\sigma_{xx}$ {\it do
not vanish at zero temperature} due to degeneracy of the Landau
levels. This correction arises when the Fermi level crosses the
Landau level.
 The limits of
kinetic equation applicability were found. We also study the
situation of kinetic memory when particles repeatedly return to
the points of their meeting.

\end{abstract}
 \maketitle
\subsection*{Introduction}
Since the discovery of quantum Hall effect the problem of 2D
electron system in strong magnetic field has attracted big
attention. It was generally accepted that the most interesting
thing is low-temperature limit when all inelastic processes are
frozen out and the system can be treated as the electron-impurity
one. Here we concentrate our consideration on the case of
semimetal with coexisting electrons and holes. Such systems based
on 2D HgTe layers were obtained and have been intensively studied
for the recent years \cite{kvon,we2009}.

The specificity of semimetal is the presence of electron-hole
scattering. Due to large density of the second component this
process can  be comparable with the impurity scattering. Usually,
in the Fermi system at $T=0$ the interparticle scattering
disappears and the friction between components gives temperature
additions $\sim T^2$ to the transport coefficients
\cite{gant,we2009}.   This is not the case in the system with a
degenerate ground state, in particular, caused by the Landau
quantization. In such a system the scattering redistributes
particles within the degenerate state that needs no energy
transfer. As a result, the friction does not disappear even at
zero temperature.

\subsection*{Problem formulation}
We consider a 2D semimetal with $g_e$ equivalent electron valleys
and $g_h$ equivalent hole valleys centered in points ${\bf
p}_{e,i}$ and centered in points ${\bf p}_{h,i}$, correspondingly.
The conduction bands with energy spectra $({\bf p}-{\bf
p}_{e,i})^2/2m_e$ overlaps with the valence bands
$E_g-\varepsilon_{{\bf p}-{\bf p}_{h,i}}$, $\varepsilon_{\bf
p}=p^2/2m_h$ ($E_g>0$). The hole mass $m_h$ is assumed  to be much
larger than the electron mass $m_e$. The distances between
electron and hole extrema $|{\bf p}_{h,i}-{\bf p}_{e,j}|$ are
supposed to be large to suppress the electron-hole conversion. At
the same time, the scattering between electrons and holes changing
the momenta near extrema are permitted. Without the loss of
generality, further we shall count the momenta from the band
extrema and replace ${\bf p}-{\bf p}_{h,j}\to {\bf p}$, ${\bf
p}-{\bf p}_{e,i}\to {\bf p}$.

The system is placed in a moderately strong magnetic field, such
that the  electrons are quantized, while holes stay classical. In
other words, the number of the filled hole Landau levels $N_h+1$
is large, while the analogical electron number $N_e+1$ has the
order of unity. We shall consider the low-temperature limit when
the electron transitions occur within the same Landau level and
transitions between different electron Landau levels are
forbidden.  The energy conservation permits this process only when
the Landau level is partially filled, i.e. in a state of
compressible Landau liquid.

We shall neglect the rearrangement of the energy spectrum caused
by the interaction between electrons and electrons and holes. The
Landau levels widening will be also neglected.

 The interaction of  quantized gas with a
classical one is an unusual situation. The kinetics of holes can
be described by a classical kinetic equation, while electrons need
a quantum description.

 The vector-potential of the magnetic field ${\bf
H}=(0,0,H)$  is chosen in the form of ${\bf A}=(0,Hx,0)$, electric
field ${\bf E}$ is directed along $x$ axis. Holes will be
described by a momentum $\bf p$ and distribution function $f_{\bf
p}$. Electrons with Landau level $n$ and momentum $k$ along $y$
axis are described by the the distribution function
$\varphi_{n,k}$.

Here we should explain some trick we will use below. If one uses
the electron states in quantizing magnetic field only, the uniform
spatial distributions of electrons gives $\varphi_{n,k}$ not
depending on $k$. Thus, there is no non-equilibrium in the
electron subsystem until one takes  off-diagonal terms of the
density matrix into consideration. In fact, the diagonal elements
of the density matrix are larger than off-diagonal ones, but they
do not determine the current. At the same time, the electron
states in the magnetic and electric fields without scattering are
localized along the field. So,  we can consider them as the first
approximation  to avoid a loss of electric field in the kinetic
equations. Hence, we shall include the electric field in the
electron states. Namely, we shall consider $|n,k>$ as the states
in the presence of the electric field directed along axis $x$.
Bearing in mind the smallness of the electric field we shall
expand the results everywhere it can be done.

In accordance with above-mentioned, we shall use the states in
crossed electric and magnetic fields $\psi_{n,k}({\bf
r})=e^{iky}\phi_n((x-X_k))/\sqrt{aL_y}$, where $\phi_n$ are
normalized oscillator functions, $L_y$ is the normalizing length
of the system in $y$-direction, $X_k=X_k^{(0)}-v_d/\omega_e, ~
X_k^{(0)}=-a^2k$ is the coordinate of the center of cyclotron
motion, ${\bf v}_d=c[{\bf E},{\bf H}]/H^2$ is the drift velocity,
$a=\sqrt{c/eH}$ is the magnetic length, $\omega_{e}=eH/m_{e}c$ is
the  electron cyclotron frequency, $-e$ is the electron charge; we
set $\hbar=1$ and will restore the dimensionality in the final
expressions. The corresponding energy is presented by
$\varepsilon_{n,k}=\omega_e(n +1/2)+eEX_k$.

\subsection*{Dissipative conductivity}
The transmission of the momentum between electrons and holes is
determined by the scattering processes. The collision term in the
kinetic equation for holes in the Born approximation reads

\begin{eqnarray}\label{20}\nonumber
    \hat{I}_{he}\{f_{\bf p}\}=\frac{2\pi}{S^2}2g_e \sum_{{\bf p}',{\bf
    q},\gamma,\gamma'}|u_{\bf q}|^2\delta_{{\bf p}',{\bf p}+{\bf
q}}|J_{\gamma',\gamma}({\bf
    q})|^2\delta(\varepsilon_{\bf p'}-\varepsilon_{\bf
    p}+\varepsilon_{\gamma'}-\varepsilon_{\gamma})\times\\\left[f_{\bf p}(1-f_{\bf
p'})\varphi_\gamma (1-\varphi_{\gamma'})-
    f_{\bf p'}(1-f_{\bf p})\varphi_{\gamma'}
    (1-\varphi_\gamma)\right].
\end{eqnarray}
Here $u_{\bf q}$ is the Fourier transform of potential of
interaction between electron and hole, $S$ is the system area,
$J_{\gamma',\gamma}({\bf    q})= <\gamma'|e^{i{\bf qr}}|\gamma>,
~~ \gamma=(n,k)$, $\gamma'=(n',k'), ~~ \varepsilon_{\bf
    p}=p^2/2m_h$ is the hole energy ($m_h$ is the hole effective mass).

Due to the uniformity of the space the quantity $\varphi_{\gamma}$
does not depend on the wave vector and coincides with the
equilibrium distribution function. At zero temperature all factors
$\varphi_{n'} (1-\varphi_{n})\equiv 0$, excluding the contribution
with $n=n'=N_e$,
     where $N_e$ is the number of the last partially filled Landau level. The
quantities $N_e$ and $\varphi_{N_e}\equiv\nu$ can be expressed via
the electron density $n_e$ as $N_e=[n_e\pi a^2/g_e], \nu=\{n_e\pi
a^2/g_e\}$ (square and figure brackets mean the integer and
fractional parts). We shall expand the collision term with respect
to weak non-equilibrium, assuming that the electric field and the
deviation of distribution function from equilibrium are small.

 Expanding Eq.(\ref{20}) we get

\begin{eqnarray}\label{22}\nonumber
   \hat{I}_{he}\{f_{\bf p}\}=\frac{2\pi}{S^2}2 g_e\sum_{{\bf
    p'},n,k}R_n(|{{\bf p}-{\bf p}'}|)\Biggl[\delta(\varepsilon_{\bf
p}-\varepsilon_{{\bf
    p}'})\Bigl[
   (\delta f_{\bf p}-\delta f_{{\bf p}'})\varphi_{n}^{(0)} (1-\varphi_{n}^{(0)})
   \Bigr]  \\+\delta'(\varepsilon_{\bf p}-\varepsilon_{\bf
    p'})eEa^2(p'_y-p_y)\varphi_{n}^{(0)} (1-\varphi_{n}^{(0)})(f_{\bf p}^{(0)}-f_{{\bf
    p}'}^{(0)})\Biggr].
\end{eqnarray}
Here $\delta f_{\bf p}$ is linear in $E$ correction to the hole
distribution function, $f_{\bf p}^{(0)},~~ \varphi_{n}^{(0)}$ are
equilibrium distribution functions of holes and electrons,
\begin{equation}\label{1111}
    R_n(q)=|u_{q}|^2(L_n^2(q^2a^2/2)e^{-q^2a^2/2},
\end{equation}
$L_n$ are the Laguerre polynomials. The function $R_n(q)$ has a
characteristic size in $q$-space $1/s$. The parameter $s$ is
determined by the largest of
 sizes of potential $L$ and wave functions of electrons
$a\sqrt{2(n+1)}$. In the coordinate space,  $s$ corresponds to the
typical impact parameter for scattering.

Let us consider the hole transport. The kinetic equation for
$\delta f_{\bf p}$ reads
\begin{equation}\label{hole eq}
e{\bf E}\frac{\partial f_{\bf p}^{(0)}}{\partial {\bf p}
}+\omega_h[{\bf p},{\bf h}]\frac{\partial \delta f_{\bf
p}}{\partial {\bf p} }=\hat{I}_{he}\{f_{\bf p}\},
\end{equation}
where  $\omega_h=eH/m_hc$,  ${\bf h}={\bf H}/H$.

The solution of Eq.(\ref{hole eq}) can be searched using a usual
substitution
\begin{equation}\label{02}
    \delta f_{\bf p}={\bf p}{\bf C}(\varepsilon_{\bf
p})\frac{\partial f_{\bf p}^{(0)}}{\partial \varepsilon_{\bf p} }.
\end{equation}
Then  Eq.(\ref{hole eq}) is algebraized
\begin{equation}\label{hole eq2}
{\bf C}+\omega_h\tau[{\bf h},{\bf C}]=-\frac{e\tau}{m_h}{\bf
E}-\frac{\tau}{\tau_{he}}{\bf v}_d\equiv-\frac{e\tau}{m_h}{\bf
E}_{ef}.
\end{equation}
Here we introduced notations:  $1/\tau=1/\tau_i+1/\tau_{he}$,
$\tau_i$ is the impurity transport relaxation time of holes. The
hole transport relaxation time,  due to hole-electron interaction,
is presented  by
\begin{equation}\label{rel time}
\frac{1}{\tau_{he}}=2g_e\frac{\nu(1-\nu)}{a^2S}\sum_{\bf
p'}\frac{({\bf p}-{\bf p'}){\bf p}}{p^2}R_{N_e}(|{{\bf p}-{\bf
p}'}|)\delta(\varepsilon_{\bf p}-\varepsilon_{{\bf
    p}'}).
\end{equation}
According to  Eq.(\ref{rel time}) the relaxation process can be
interpreted as energy-conserving collisions of  holes with
immobile distributed charges with form-factors $R_N(q)$ and
density $\nu(1-\nu)/\pi a^2$. The factor $\nu(1-\nu)$ reflects the
circumstance that the scattering occurs due to the fluctuations of
electron density; fully occupied Landau states produce no
fluctuations. One can introduce the  quantum cross-section of
scattering $\sigma$   by the relation $\tau_{he}=\pi a^2/\sigma
\nu(1-\nu)v_{Fh}~~(v_{Fh}$ is the hole Fermi velocity). The
quantity $\sigma$, unlike $s$, characterizes the strength  of
scattering, rather than the size of the scatterer.

After simplifications, \begin{equation}\label{rel time2}
\frac{1}{\tau_{he}}=2g_e\nu(1-\nu)\frac{m_h}{2\pi^2p^2a^4}\int_0^{2p^2a^2}
e^{-t}L_{N_e}^2(t)~|u_{\sqrt{2t}/a}|^2~\sqrt{\frac{t}{2p^2a^2-t}}dt.
\end{equation}

The effective electric field consists of the external electric
field (the first contribution in the middle expression in
Eq.(\ref{hole eq2})) and the drag force from electrons, drifting
across the external field (the second term).

Solving the equation (\ref{hole eq2}) we get
\begin{equation}\label{A}
    {\bf C}=-\frac{e\tau}{m_h} \frac{{\bf E}_{ef}-\omega_h\tau[{\bf h},
    {\bf E}_{ef}]}{1+(\omega_h\tau)^2},
    \end{equation}
The hole current at temperature $T=0$ is

    \begin{equation}\label{jh}
    {\bf j}_h=\frac{e^2n_h\tau}{m_h}
    \frac{1}{1+\omega_h^2\tau^2}
    \left[(1-\frac{\tau}{\tau_{he}}){\bf
E}-(\frac{1}{\omega_h\tau_{he}}+\omega_h\tau)[{\bf h
    E}]\right],
    \end{equation}
    where $n_h$ is the hole concentration, relaxation parameters
$\tau_{he}$ and $\tau$ are taken at the Fermi energy of holes.

For determination of the dissipative electron current we shall use
the Titeika formula \cite{titeika, adams})
\begin{equation}\label{tit}
   j_{e,x}=2g_e\frac{e}{S}\sum_{n,k',k}(X_{k'}-X_k)W_{n,k';n,k}\varphi_{n,k'}
    (1-\varphi_{n,k}),
\end{equation}
where $W_{\gamma',\gamma}$ is the electron transition probability
from $\gamma'$ to $\gamma$:
\begin{eqnarray}\label{21}\nonumber
    W_{\gamma',\gamma}=4g_h\frac{\pi}{S^2}\sum_{{\bf p},{\bf p}',{\bf
    q}}|v_{\bf q}|^2\delta_{{\bf p}',{\bf p}+{\bf q}}|J_{\gamma',\gamma}({\bf
    q})|^2\delta(\varepsilon_{\bf p}-\varepsilon_{\bf
    p'}+\varepsilon_{\gamma'}-\varepsilon_{\gamma}) f_{\bf p'}(1-f_{\bf
    p}).
\end{eqnarray}
 The physical meaning of the formula (\ref{tit}) is obvious: the
scattering of an electron  changes  its coordinate of cyclotron
center $X_k$. After expansion we find
\begin{eqnarray}\nonumber
W_{n,k';n,k}\varphi_{n,k'}
    (1-\varphi_{n,k})\approx \frac{4\pi g_h}{S^2}\sum_{{\bf p},{\bf
    p'}}\delta_{k'-k,p'_y-p_y}R_n(|{{\bf p}-{\bf p}'}|)(F_1+F_2),
    \\F_1=
   \delta(\varepsilon_{\bf p}-\varepsilon_{{\bf
    p}'})[(\delta f_{{\bf p}}(1-f_{\bf p'}^{(0)})-\delta f_{\bf p'}f_{\bf
p}^{(0)})\varphi_{n}^{(0)}
    (1-\varphi_{n}^{(0)})],\nonumber\\F_2=
        -eEa^2(p'_y-p_y)\delta'(\varepsilon_{\bf p}-\varepsilon_{{\bf p'}})f_{\bf
p'}^{(0)}(1-f_{\bf p}^{(0)})\varphi_{n}^{(0)}
    (1-\varphi_{n}^{(0)}).\label{233}
    \end{eqnarray}
The contribution  $F_1$ is connected with the absence of
equilibrium in the hole subsystem. It could be calculated with
using  the electron states at ${\bf E}=0$. The electric field
enters
 this contribution indirectly, via $\delta f_{\bf p}$. The term
$F_2$ is determined by the direct perturbation of electronic
states by the electric field. These contributions are independent
and appear or do not appear in different kinetic problems.

 Generally speaking, the first contribution to the current is not bound to the
direction of the electric field. The specific  choice of gauge
${\bf A}=(0,Hx,0)$ gives possibility to calculate only $x$
component of this current. The $y$ component of this current can
be found owing to the fact that the resulting current is
independent on the gauge,  by means of the choice ${\bf
A}=(-Hy,0,0)$. Collecting $j_x$ and $j_y$ components we find  the
electron current density caused by the nonequilibrium of holes
\begin{equation}\label{jdf}
  {\bf j}^{(1)}_e=2g_eg_h\frac{e}{S^2}\nu(1-\nu)\sum_{{\bf p},{\bf p}'}~
  \big[({\bf p}'-{\bf p})
~  {\bf h}\big]\delta(\varepsilon_{\bf p'}-\varepsilon_{\bf
p})~R_{N_e}(|{\bf p}'-{\bf p}|)~(\delta f_{\bf p'}-\delta f_{\bf
p})
\end{equation}
Substituting  (\ref{02}) and (\ref{A}) into Eq.(\ref{jdf}) we get
to
\begin{equation}\label{}
  {\bf j}^{(1)}_e=\frac{1}{\omega_h\tau_{he}}[{\bf j}_h{\bf
  h}].
\end{equation}
 For the dissipative current
caused by  the direct action of electric field we find:
\begin{equation}\label{jE}
    {\bf j}^{(2)}_e=2g_eg_h\frac{e^2a^2}{2S^2}\nu(1-\nu){\bf E}\sum_{{\bf p},{\bf p}'}~
  \delta(\varepsilon_{\bf p'}-\varepsilon_{\bf
p})~\delta(\varepsilon_{\bf p'}-\varepsilon_{F,h})~({\bf p}'-{\bf
p})^2~R_{N_e}(|{\bf p}'-{\bf p}|).
\end{equation}
Here $\varepsilon_{F,h}$ is the Fermi energy of holes.

Taking into account Eq.(\ref{rel time}) one can reduce
Eq.(\ref{jE}) to the form
\begin{equation}\label{jE2}
    {\bf j}^{(2)}_e=\frac{n_he^2}{m_h}\frac{1}{\omega_h^2\tau_{he}}{\bf E}
\end{equation}

Using  the preceding formulas  we arrive at the expressions for
the  dissipative  conductivity of the system with interacting
electrons and holes
\begin{eqnarray}\label{ff}
\sigma_{xx}=n_h\pi a^2\frac{e^2}{\pi
\hbar}~~\frac{1-\tau/\tau_{he}}{\tau_{he}\omega_h}~~\frac{1+\tau\tau_{he}\omega_h^2}{1+\tau^2\omega_h^2}.
\end{eqnarray}
Here $n_h\pi a^2/g_h\approx N_h+1$ is a large number.  The
dissipative conductivity Eq.(\ref{ff}) vanishes if $1/\tau_i=0$.
Indeed, in the absence of impurities the Lorentz invariance
permits to exclude electric field  by the transition to the frame
of reference drifting across the electric field with the velocity
${\bf v}_d$. The evident absence of the energy losses in this
frame  proves the absence of $\sigma_{xx}$.

It is useful to subtract from  (\ref{ff}) the trivial contribution
of holes in the absence of electron-hole scattering
$\sigma_{xx}|_{\tau_{he}\to\infty}$ which corresponds to $\nu=0$
and can be easily found experimentally in the incompressible
phase.
 The result is
\begin{eqnarray}\label{ffex}\delta\sigma_{xx}=n_h\pi a^2\frac{e^2}{\pi\hbar}~~\frac{\tau_{he} + \tau_i -
\tau_{he} \tau_i^2 \omega_h^2}{\omega_h (1 + \tau_i^2 \omega_h^2)
((\tau_{he} + \tau_i)^2 + \tau_{he} ^2 \tau_i^2
  \omega_h^2)}\end{eqnarray}
  Though the searched effect is caused by the
friction between components it vanishes if $\tau_{he}\to\infty$ or
$\tau_{i}\to\infty$. The applicability of Eq.(\ref{ff}) is limited
by the large magnetic field because $H$ should quantize electrons.

The figure shows the correction to the dissipative conductivity
$\delta\sigma_{xx}$ conductivity in units of conductance quantum
$e^2/\pi\hbar$ {\it versus} the inverse magnetic field in the
neutrality point $ n_e=n_h$. We utilized the screened Coulomb
potential $u_q=2\pi e^2/(\chi(q+\varkappa))$, where $\chi$ is the
dielectric constant. For calculations we approximate  the
screening parameter $\varkappa$ by the value $2/a_{B,h}$ ~
($a_{B,h}$ is the Bohr radius for holes). The parameters of HgTe
from \cite{we2009} $\chi = 20, \ m_e = 0.025 m_0,\  m_h = 0.15
m_0, \ g_e=1, \ g_h=2$ have been taken. For the energy gap we have
chosen the value $ Eg = 0.003$ eV. The quantity
$\delta\sigma_{xx}$ is finite when the Fermi level gets to the
Landau level.
\begin{figure}[h!]\label{fig}
\centerline{\epsfxsize=10cm \epsfbox{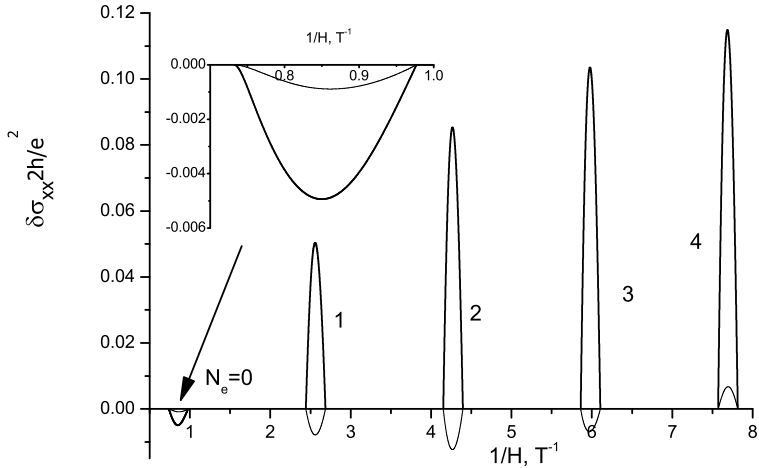} }
 \caption{Correction to the dissipative conductivity $\delta\sigma_{xx}$ due to e-h scattering {\it versus} the
magnetic field for $\tau_i = 10^{-12}$s (thick line) and for
$\tau_i = 3\cdot10^{-12}$s (thin line). }\end{figure}

\subsection*{Finite Landau level width}
The crucial point for previous consideration, in particular,  for
temperature-independent contribution of e-h scattering to the
dissipative conductivity is the presence of the Landau levels
degeneracy. The potential fluctuations lift this degeneracy and
the the Landau levels get width $\gamma_n$. For long-range
potential this width is  proportional to the amplitude of
potential. For developed fluctuations the self consistent Born
approximation \cite{ando} gives the width of the Landau levels
proportional to $\sqrt{n_i}$ ($n_i$ is the impurity concentration)
and the potential of individual impurity. The exception is the
short-range impurities with $\delta$-like potential for which the
part of the Landau level states $1/\pi a^2-n_i$ remains degenerate
if $1/\pi a^2>n_i$, while  $n_i$ states form a band of localized
with a finite width \cite{we1978}.

In the case of the Landau level with a finite width $\gamma_N$ the
interparticle scattering  depends on the ratio of the temperature
$T$ to the width. If  $T\ll \gamma_n$ the e-h scattering is
suppressed and if $T\gg \gamma_n$ the scattering does not notice
the width. Thus, all previous consideration of e-h scattering
remains valid for intermediate temperature
$\hbar\omega_e>T>\gamma_n$.

The scattering processes work in the dissipative conductivity of
quantized electrons in a parallel manner. One can sum up the
contributions to electron $\sigma_{xx}$ conductivity caused by
impurity  scattering  and electron-hole processes. According
\cite{ando}, the  contribution to electron dissipative current
caused by short-range impurity scattering is
\begin{equation}\label{ando}
    (\sigma_{xx})_{ei}=\frac{g_ee^2}{\pi^2\hbar}(N_e+1/2)(1-\mu^2).
\end{equation}
The reduced distance between the Fermi level and the Landau level
with number  $N_e$,
$\mu=(\varepsilon_{F,e}-(N_e+1/2)\omega_e)/\gamma_N,$ is connected
with the quantity  $\nu$ by the equation
$$\nu=\frac{1}{2\pi}(\pi+2\mu\sqrt{1-\mu^2}+2\arcsin\mu).$$

 The result Eq.(\ref{ando}) should be added to (\ref{jE2}). The ratio of the
 contribution to conductivity
 (\ref{jE2}) to (\ref{ando}) in their maxima is
\begin{equation}\label{ei}
    \frac{\pi g_h
 (N_h+1)}{g_e(N_e+0.5)}\frac{1}{\omega_h\tau_{he}}.
\end{equation}
 The first factor is large by assumptions, the second one can be
 as large, so small.
 The total dissipative conductivity is the sum of (\ref{ff}) and
 (\ref{ando}).

 In the case of short-range impurities with  small concentration $n_i\ll 1/\pi
a^2$  the absence of widening leads to the validity of the results
obtained in  the previous section up to zero temperature. In more
wide range $n_i< 1/\pi a^2$ the scattering rate $1/\tau_{he}$
should be corrected by the factor $ 1-n_i\pi a^2$ reflecting the
fraction of degenerate states.

If the long-range potential fluctuation case is realized the
degeneracy disappears. In the absence of e-h scattering the model
of adiabatic transport is valid when electron cyclotron centers
are drifting along the lines of constant potential. Without the
external field only one infinite  fractal level line of the
fluctuating potential  corresponding to the percolation threshold
exists. In the presence of the finite electric field this level
line decays to independent infinite entangled lines going across
the external electric field. The drift does not depend on the
charge of particles:  the velocities and trajectories of the
cyclotron centers of quantized electrons and classical holes are
the same. The dissipative conductivity of electrons vanishes,
while the Hall conductivity changes stepwise between quantized
$Ne^2/h$ values. In the lack of
 degeneracy the temperature-independent e-h scattering also
 disappears.

 The dissipative conductivity could result from the jumps
 between  different energy levels with the transfer of abundant energy between
electrons
 and holes. However, electrons and holes exactly repeat
 trajectories of each other and corresponding contributions
 cancel.

\subsection*{Memory effects} Here we discuss the memory effects caused by the
scattering between components. In the previous sections the
validity of the kinetic equation for holes was assumed. The
presence of strong enough classical magnetic field violates the
applicability of the kinetic equation.  The inapplicability is
reasoned by the lack of randomness of  motion  of a charged
particle due to recurrent returns to the starting point. Let the
scatterer has a finite size. The particle either never meets a
scatterer or returns to the scatterer if the particle has met the
scatterer once.

 Starting from the paper \cite{we1978}, these "memory" effects were subject of
investigation in relation to the impurity scattering   (see also
more recent  papers\cite{fogl,mirl1,we1998,mirl2,dmit}). In a
strong magnetic field the picture of classical transport  of holes
interacting with quantized electrons is similar to the 2D hole
system with short-range impurities. The inelasticity of scattering
is absent due to the degeneracy of the Landau level and electrons
play a role of immobile scatterers for holes. Their effective
density is $g_e\nu (1-\nu)/\pi a^2$. The parameter $x=g_e\nu
(1-\nu)$ is the main factor governing the number of these
scatterers. The quantity $s$ serves as a spatial size of
scatterers. If $s$ is less than the cyclotron radius, the
electrons can be treated as short-range scatterers for holes. The
total density of scatterers $\tilde{n}$ is the sum of the impurity
$n_i$ and active electron densities $\tilde{n}=n_i+x/\pi a^2$.

The kinetic equation is applicable  if  holes experience
 multiple collisions   on their cyclotron path
$2\pi r_h=2\pi\sqrt{2 (N_h+1)}a$, where $N_h \gg 1$ is the number
 of the last-filled hole Landau level. In other words, the
 quantity
$l=1/(2s\tilde{n})$ should be less than $2\pi r_h$. Assuming that
$s\sim a\sqrt{2(N_e+1)}$ we get the condition
\begin{equation}\label{cond}
   8(\pi n_ia^2+x)\sqrt{(N_e+1)(N_h+1)}\gg 1.
\end{equation}
If $8\pi n_ia^2\sqrt{(N_e+1)(N_h+1)}\gg 1$ the kinetic equation is
applicable for any $\nu$. Otherwise,  the kinetic equation is
valid in the domain $8x\sqrt{(N_e+1)(N_h+1)}\gg 1$ that covers
(since $N_h\gg 1$) the majority of the range of variation of
$\nu$, including the center of the Landau level $\nu= 0.5$. The
applicability of the kinetic equation is violated on the wings of
the Landau level, where $\nu\ll 1$ or $1-\nu\ll 1$.

If the density of scatterers is small, $8(\pi
n_ia^2+x)\sqrt{(N_e+1)(N_h+1)}\ll 1$, the most part of holes,
namely $\exp(-2\pi r_h/l)$, will not collide the scatterers at
all. The remainder $1-\exp(-2\pi r_h/l)$ collides the scatterers.
The criterium (\ref{cond}) differs from the criterium of a strong
field from the kinetic equation point of view $\omega_h\tau\gg 1$.
Even if to suppose the equality of cross-section $\sigma$ and
$2s$, the inequality $\omega_h\tau\gg 1$ gives $r_h\ll l$, that
differs from $2\pi r_h\ll l$ by the numerical factor $2\pi$.

If $\tilde{n}\pi r_h^2\ll 1$ the particle being scattered once,
returns the scatterer repeatedly. The trajectory of a hole
consists of circular segments composing, in a general case, an
open aperiodic rosette. During  a long time the trajectory covers
a circle of radius approximately equal to $2r_h$. Such
trajectories are localized and $\sigma_{xx}=0$.

This picture is valid if there are no other scatterers inside this
circle. Otherwise, the rosettes around two or more scatterers join
together. When the scatterers whose distance from each other is
less than $2r_h$ compose a infinite chain, the rosettes join an
infinite cluster, the hole gets possibility to travel on the
infinite distance and the finite conductivity $\sigma_{xx}$
appears.

 An accurate criterium of localization  is given by the theory of
 percolation. According to \cite{shkl,Essam1,Essam2}, the percolation threshold
reads $\tilde{n}\pi r_h^2=B_c$, where $B_c$ is  some number.
 If to supposed that all impurities and electrons are randomly distributed the
corresponding mathematical problem is the percolation threshold
for the uniform distribution of disks of radii $r_h$, where
 $B_c=1.128$. The critical value of $x$ is
$x_c=B_c/2(N_h+1)-\pi n_i a^2$, assuming that $2\pi n_i
a^2(N_h+1)<B_c$.
 Since $N_h\gg1$, the
critical values of $\nu$ or $1-\nu$ are small: the holes become
localized when the Fermi level gets to the wings of  electron
Landau levels.

Above the threshold (if $2\pi n_i a^2(N_h+1)>B_c$ or $x>x_c$) the
rosettes overlap and the diffusion chain is formed: a particle
executes a motion around one scatterer, then it jumps to a rosette
around the other, etc.. If $2\pi n_i a^2(N_h+1)<B_c$ and $x_c<x\ll
(8\pi\sqrt{(N_e+1)(N_h+1)})^{-1}$ the non-colliding trajectories
of holes are most frequent, but the colliding trajectories form
the infinite cluster.

Near the threshold the conductivity in the conductive phase can be
estimated. The parameter $x-x_c\ll x_c$  characterizes  the
proximity with the threshold. Only a small fraction of holes $2\pi
r_h/l$ can collide the scatterers. This fraction can form the
rosettes, while others play a passive role. The holes colliding
with electrons have a mean free path of $2\pi r_h$.

Let all scatterers be situated in a lattice with a period less
than $2r_h$. In this case the rosettes around scatterers unite and
the conductivity can be estimated as a conductance of the
elementary cell
$$\sigma_0=\frac{e^2\tau_{he}}{m_h} n_h\frac{2\pi r_h}{l}.$$
If the scatterers are randomly distributed  the percolational
cluster near the threshold of its formation is rarefied. Hence,
the conductivity of the system is less than $\sigma_0$. Over the
distance larger than $2r_h$ the medium can be treated as a
continuum mixture of conducting cells with conductivity $\sigma_0$
and insulating cells. The effective conductivity can be found
using the  critical index for 2D conductivity $t$:
$\sigma_{xx}=\sigma_0 (x-x_c)^t$, $t=1.2$.

Summarizing, the hole conductivity has the symmetric dependence on
the quantity $0<\nu<1$ via the factor $x=\nu(1-\nu)$ and the
maximum at $\nu=0.5$.  The conductivity critically vanishes on the
far wings $x<x_c\ll 1$. Near the threshold ($x-x_c\ll x_c$) the
conductivity has a power-like growth. Farther from the edges the
domains of the coexistence of localized circular motion and
diffusion on combined infinite rosettes are situated.  The central
region corresponds to the usual transport described by the kinetic
equation.

\subsection*{Conclusions}
We have studied the influence of electron-hole interaction on
transport in the system where electrons are quantized and holes
are not. In these conditions, the  second type of carriers plays
its role as an additional (or exceptional) channel of scattering.
Weak electron-hole interaction can be considered in the Born
approximation, despite the degeneracy of the Landau levels, in
contrast to the impurity mechanism which is not perturbative in
the quantizing magnetic field, even for a weak potential. The
scattering of holes on quantized non-interacting electrons occurs
if, and only if,  the Landau level is partially filled. The
chaotization results from the random distribution of electrons in
the momentum space and corresponding entropy and is not frozen out
at zero temperature and remains finite. The scattering of holes
can be considered by means of kinetic equation approximation when
the Fermi level is near the center of the Landau levels;  the
kinetic approximation loses applicability apart from the center
and on the far wings the holes become localized.

\end{document}